\documentclass[12pt]{article}
\newcommand{\farcs}{${''}$\nolinebreak\hspace{-1.4mm}.}
\newcommand{\cthead}[1]{\multicolumn{1}{|c|}{#1}}
\usepackage{graphicx,afterpage}
\begin{document}
\begin{center}
\Large\bf
METHANOL AND H$_2$O MASERS IN A DISK AROUND GL~2789
\end{center}
\begin{center}
{\Large
V.~I.~Slysh$^1$, M.~A.~Voronkov$^1$, I.~E.~Val'tts$^1$, V.~Migenes$^2$}
\parbox{0.85\linewidth}{
\begin{itemize}
\item[$^1$~-] Astro Space Center, Lebedev Physical Institute, 84/32
Profsoyuznaya ul., 117997 Moscow, Russia
\item[$^2$~-] Departamento de Astronom\'ia, Universidad de Guanjuato,
Guanjuato, Mexico
\end{itemize}}\\
Received on March 20, 2002; In final form on May 23, 2002
\end{center}

\begin{abstract}
VLBA and EVN radio observations of H$_2$O masers at 22~GHz and methanol
masers at 6.7~GHz have been used to obtain images of the maser spots
in the infrared object GL~2789, which is associated with the young stellar
object V645~Cyg. The position of these masers coincides with that of the
optical object to within 0\farcs2. The maser spots are located
in a line oriented north--south, and their positions and radial velocities
can be described by a model with a Keplerian disk with maximum radius
40~AU for the H$_2$O masers and 800~AU for the methanol masers. The H$_2$O
and methanol masers spots are unresolved, and the lower limits for their
brightness temperatures are $2\times10^{13}$~K and $1.4\times10^9$~K,
respectively. A model in which the maser radiation is formed in extended
water--methanol clouds associated with ice planets forming around the young
star is proposed.
\end{abstract}

\section{INTRODUCTION}
GL~2789 is an infrared source that coincides with an optical reflection
nebula and the star-like object V645~Cyg [1]. The optical image contains
the star-like condensation N0 and several filamentary nebulae, the brightest
of which is called N1. Cohen~[1] suggests that this object is associated
with a young O7 star at a distance of 6~kpc that is surrounded by a bipolar
nebula. Humphreys et al.~[2] concluded that the optical spectrum of the
nebulosity around V645~Cyg corresponded to that of an A star surrounded
by a shell, and estimated a smaller distance to the object of about
3.5~kpc (see also Goodrich~[3]). In both cases, the mass of the star was
estimated to be 10~$M_\odot$. V645~Cyg is located at the center of a
molecular cloud that emits CO and NH$_3$ lines, with the CO line emission
showing signs of a bipolar outflow~[4, 5]. The optical object V645~Cyg is
also located at the center of a source of thermal radio continuum emission
about $7''$ in size~[6]. Lada et al.~[7] discovered H$_2$O masers with
radial velocities of ${-48.9}$~km/s and ${-44.5}$~km/s, whose position
was determined with an accuracy of 0\farcs2 and coincides with
the position of the optical component N0. The VLA observations of
Tofani et al.~[8] obtained 13 years later showed H$_2$O masers with radial
velocities of ${-43.3}$~km/s and ${-41.0}$~km/s at the same position.
OH maser emission at 1665~MHz at radial velocities ranging from ${-45}$~km/s
to ${-41.6}$~km/s was detected by Morris and Kaz\'es~[9]. Slysh et al.~[10]
detected a class~II maser in the $5_1-6_0A^+$ methanol line at
6.7~GHz at radial velocities from ${-43.5}$~km/s to ${-40.5}$~km/s. Since
maser emission is associated with early stages of stellar evolution, the
presence of OH, H$_2$O, and methonol masers suggests that V645~Cyg is
a protostar or young stellar object. Investigations of the fine structure
of the maser sources could shed light on mechanisms for interactions between
the young star or protostar and its surrounding medium. Here, we present
the results of high-angular-resolution (milliarcsecond-scale) studies
of the structure of the H$_2$O and methanol masers associated with this
source.

\section{OBSERVATIONS}

\begin{figure*}[t!]
\includegraphics[width=\linewidth]{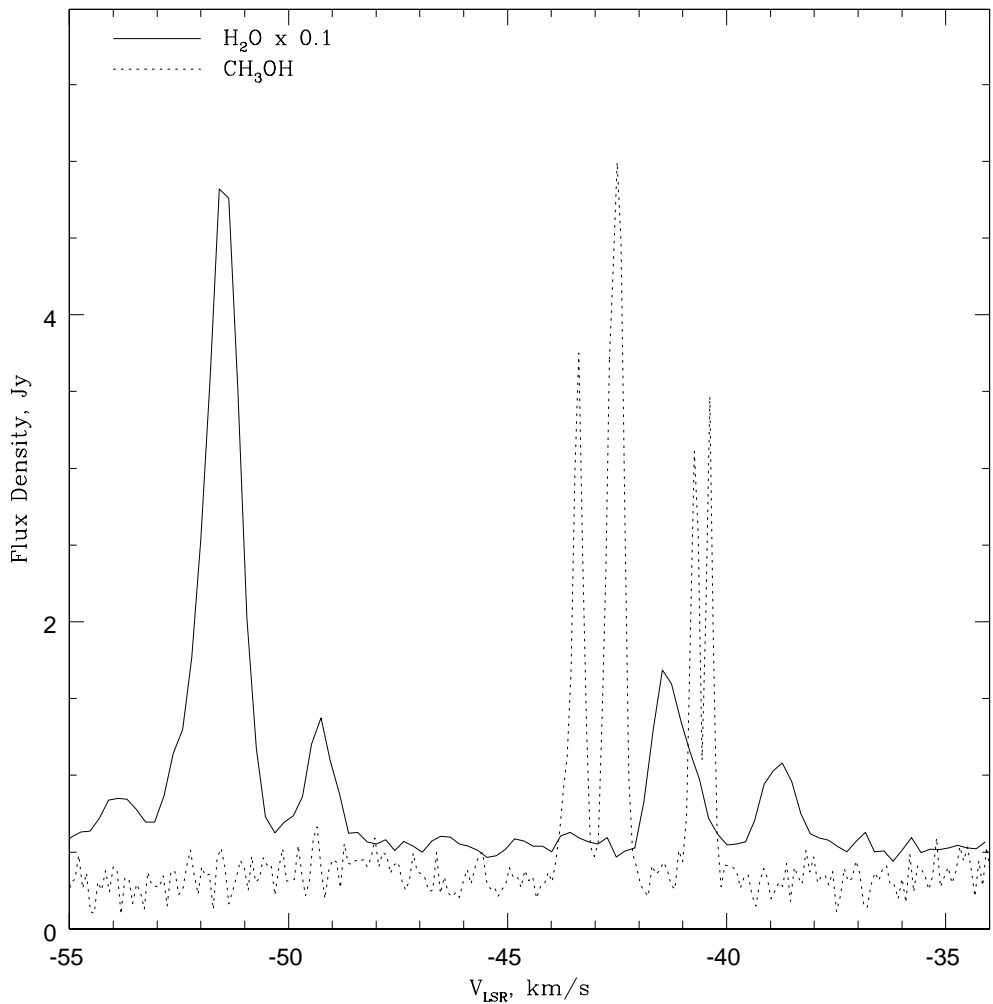}
\caption{Water vapor (solid) and methanol (dotted) maser spectra of GL~2789.
The CH$_3$OH maser spectrum was constructed from observations obtained
in 1998. The flux density of the H$_2$O maser has been decreased by a
factor of ten. \hfill}
\end{figure*}

The H$_2$O-maser observations of GL~2789 were carried out at 22~GHz on the
VLBA on June 5, 1996 as part of a prelaunch survey for the VSOP space-VLBI
project~[11]. The VLBA consists of ten 25-m radio telescopes located at
various points on US territory, and provides a maximum baseline length of
about 8000~km. The observations were conducted in a snapshot mode with
the individual scans being five minutes in duration. The resulting synthesized
beam was 0.3$\times$0.95~mas. The total recorded bandwidth of 8~MHz was
divided into 512 spectral channels during the correlation of the data,
providing a spectral resolution of 15.6~kHz per channel or 0.21~km/s.
The sensitivity achieved during the five-minute accumulation time was
100~mJy/beam. The delay and bandpass calibration was performed using
observations of the continuum sources 3C~273 and 3C~84. The post-correlation
reduction included amplitude calibration and fringe fitting. The fringe
fitting was carried out in two stages: (1) the residual delays for
the radio telescopes were determined using the continuum sources and (2)
the residual fringe rate was determined for each of the sources
individually using its strongest spectral features. We also applied a
correction for the Doppler shift of each telescope.

Observations of the class~II methonol maser in the $5_1-6_0A^+$
transition were carried out at 6.7~GHz on the European VLBI network
(EVN) in 1998 and 2000. Five telescopes equiped with 6.7-GHz receivers
participated: the 100~m Effelsberg, 25~m Jodrell Bank, 32~m Medicina (Noto
in 2000), 25~m Onsala, and 32~m Torun antennas. In 1998 and 2000,
GL~2789 was observed
during four and three 10-min scans, respectively. The resulting synthesized
beams were $4.3\times8.2$~mas. The recorded bandwidth of 2~MHz was divided
into 1024 spectral channels, providing a spectral resolution of 1.95~kHz
or 0.088~km/s. The sensitivity of the observations was 400~mJy/beam.
The delay and bandpass calibration was conducted using observations of
several continuum sources. The post-correlation reduction of the methanol
observations was carried out in the same way as the reduction of the
H$_2$O-maser data.

The subsequent processing of the data consisted of determining the absolute
coordinates of the masers and the relative positions of the maser spots, and
constructing images of the spots themselves. This was carried out in the
AIPS package.

\section{RESULTS}

\subsection{Spectra}

\begin{figure*}[t!]
\includegraphics[width=\linewidth]{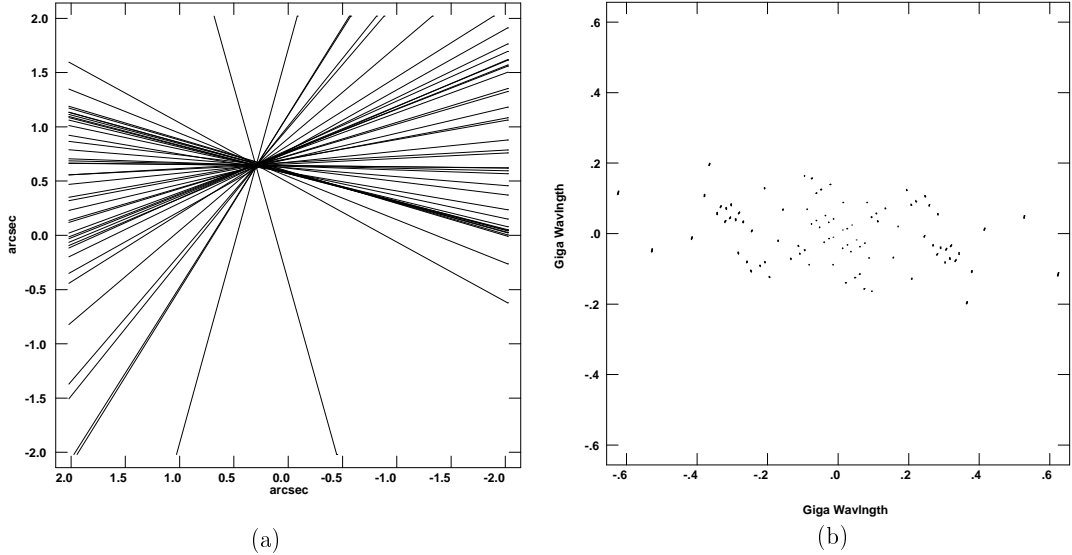}
\caption{(a) Determination of the absolute coordinates of the H$_2$O maser
($v=-51.8$~km/s). The shift of the intersection point from the phase center
(0,0) is $\Delta \alpha=293.9\pm11.6$ mas, $\Delta \delta =650.8\pm3.7$ mas,
and corresponds to the difference between the true maser coordinates at
epoch and the coordinates used during correlation. (b) $uv$ coverage for
the observations of the H$_2$O maser in GL~2789.\hfill}
\end{figure*}

Figure~1 presents the spectra of the H$_2$O (solid) and methonol (dotted)
masers observed in 1998. Each spectrum contains four features, however the
H$_2$O maser occupies a broader velocity interval than the methanol maser.
The entire methanol spectrum is located in the red half of the
H$_2$O radial-velocity spectrum. No methanol maser feature coincides in
radial velocity with any of the H$_2$O maser features. A comparison with
the earlier observations of Lada et al. ~[7] and Tofani et al.~[8] shows
that the H$_2$O spectra experienced substantial variations. In 1998, the
brightest feature had a radial velocity of ${-51.5}$~km/s, while the
brightest features in 1979~[7] and 1992~[8] had radial velocities of
${-50.2}$~km/s and ${-43.3}$~km/s, respectively. Features at radial velocities
of ${-39}$~km/s and ${-41.5}$~km/s have also appeared. The methanol maser
spectrum for the 1998 EVN observations differs markedly from a spectrum
observed at Medicina in 1995, but nearly coincides with the spectrum obtained
by Szymczak et al.~[12] in 1999. While the brightest feature in 1995 had
a radial velocity of ${-43.6}$~km/s and a flux of 19~Jy, the brightest features
in 1998, 1999, and 2000 were at radial velocities of ${-42.5}$~km/s (flux
7.5~Jy), ${-40.9}$~km/s (flux 7~Jy), and ${-40.4}$~km/s (flux 27~Jy),
respectively. At the same time, the character of the spectrum remained
nearly unchanged, always containing the same four features, with only
their relative fluxes varying.

\subsection{Absolute coordinates}

We determined the absolute coordinates of the masers using the method
of fringe-rate mapping. Figure~2a shows the result of measuring the
absolute coordinates of the H$_2$O feature at ${-51.8}$~km/s. Each line
corresponds to the locus of possible source positions for which the
fringe rate for one of the VLBA baselines is equal to the measured value.
The number of lines is equal to the number of baselines (the number of pairs
of antennas). Since the VLBA consists of ten antennas, the number of
baselines is 45. The point of intersection of all the lines corresponds to
the true position of the source. The accuracy of the absolute coordinates
of H$_2$O masers obtained in this way varies from fractions of an arcsecond
to several milliarcseconds. The coordinates of the methanol masers were
determined in the same fashion.

\begin{table*}[t!]
\caption{Coordinates of the brightest H$_2$O and CH$_3$OH maser features in
GL~2789}
\scriptsize
\vskip 3mm
\begin{tabular}{l|l|l|l}
\hline
\multicolumn{1}{c|}{Measurement}&\multicolumn{1}{c|}{{$\alpha_{2000}$}}&\multicolumn{1}{c|}{{$\Delta _{2000}$}}&\multicolumn{1}{c}{Reference}\\
\hline
Optical, N0                        &  $21^h39^m58^s\hspace{-1.35mm}.27\pm0^s\hspace{-1.35mm}.03$& $50^{\circ}14'20''\hspace{-1.4mm}.9\pm0''\hspace{-1.4mm}.2$  & Cohen [1] \\
Radio, 3.6 and 6 cm, VLA         & $ 21~~39~~58.26\pm0.05$& $50~~14~~21.3\pm0.5$      & Skinner et al. [6]\\
H$_2$O maser, VLA ($-43.3$~km/s)  &  $21~~39~~58.27\pm0.01$& $50~~14~~21.0\pm0.1$       & Tofani et al. [8] \\
H$_2$O maser, VLBI ($-50.2$~km/s) &  $21~~39~~58.28\pm0.03$& $50~~14~~21.0\pm0.2$      & Lada et al. [7] \\
H$_2$O maser, VLBA ($-51.8$~km/s) &  $21~~39~~58.277\pm0.001$ & $50~~14~~21.041\pm0.005 $ & Current paper   \\
Methanol, EVN 1998 ($-42.5$~km/s) &  $21~~39~~58.286\pm0.006$ & $50~~14~~20.98\pm0.05$   & Current paper  \\
Methanol, EVN 2000 ($-42.5$~km/s) &  $21~~39~~58.286\pm0.005$ & $50~~14~~20.6\pm0.2$   & Current paper  \\
\hline
\end{tabular}
\end{table*}

Table 1 presents the coordinates of the brightest maser features. The table
presents the coordinates obtained from both our 1998 and 2000 methanol
observations, which agree within their errors ($<2\sigma$). The coordinates
of the H$_2$O maser coincide with those presented by Lada
et al.~[7] and Tofani et al.~[8] within their errors, although they
correspond to different spectral features. The accuracy for the H$_2$O
maser coordinates provided by our observations was more than an order of
magnitude higher than for the observations of Lada et al.~[7] and Tofani
et al.~[8]. The methanol maser is shifted relative to the H$_2$O
maser by 0\farcs08$\pm$0\farcs07 to the east and
0\farcs09$\pm$0\farcs05 to the south. The coordinates of
both masers coincide with those for the optical object N0 and the source
of continuum radio emission, which are known with much lower accuracy.
Nevertheless, we can assert with certainty that the masers are located
within the extended continuum radio source, which has a diameter of~7$''$.

\subsection{The Images}
\subsubsection*{3.3.1. The H$_2$O maser}
To construct the image of the
H$_2$O maser, we performed self-calibration of the VLBA data using the strongest
spectral feature, at a radial velocity of ${-51.8}$~km/s. We were not able
to calibrate using external sources, i.e., strong point-like continuum
sources, since the amplitude and phase characteristics of the array did not
remain sufficiently constant between observations of the calibrator
and maser sources, due to changing conditions in the atmosphere. During
the self-calibration on a maser feature, the amplitude and phase
characteristics of the antennas were determined simultaneously with the
image of the maser itself, and variations in these characteristics and
in the atmospheric characteristics were removed from the data.

Figure~2b shows the coverage of the $uv$ plane for the observations of
GL~2789. Since each of the individual observations was short (5~min), the
tracks corresponding to individual pairs of antennas are represented by
short segments of ellipses that appear nearly like dots. We can see that
the range for $u$ (horizontal direction) is appreciably larger than the
range for $v$ (vertical direction). This is due to the locations of
the antennas in the VLBA, whose extent in the east--west direction is
somewhat larger than in the north--south direction. As a result, the
synthesized beam has an elliptical shape with a 3:1 axial ratio.

The self-calibration procedure yielded an image of the reference feature
at ${-51.8}$~km/s, as well as images of all the remaining features with
radial velocities from ${-38.0}$~km/s to ${-55}$~km/s. The maser spots
are concentrated primarily in two groups. The southern group contains
features with radial velocities from ${-49}$~km/s to ${-55}$~km/s, while
radial velocities in the northern group are from ${-38}$~km/s to ${-42}$~km/s.
Figures~3 and 4 present images of the maser spots of the northern and southern
groups in all channels where there was an appreciable signal. These images
are super-resolved; i.e., the images were convolved with a beam 0.1~mas
in size. This is admissible in the presence of high signal to noise. A
map of the masers in the southern group constructed for all the
channels is shown in Fig.~5a, an analogous map for the northern group is
shown in Fig.~5b, and a combined map is shown in Fig.~5c. We can see that
the southern group of maser spots forms a structure that is elongated
in the east--west direction, while the northern group is elongated roughly
north--south. Within the northern group, the radial velocity decreases
from south to north, contrary to the general tendency for the radial
velocity for the maser as a whole to grow from south to north. One weak
feature with a radial velocity of ${-50.5}$~km/s has a peculiar location,
separated from the southern group by about 5.6~mas to the southwest. In all,
we detected eight maser spots in the image of GL~2789. Their mutual
locations relative to the reference feature at ${-51.8}$~km/s are presented
in Table~2 and shown in Fig.~5d. This maser source was the most compact of all
those observed during the survey: the maximum distance between individual
features does not exceed 10~mas.

\begin{figure*}[t!]
\includegraphics[width=\linewidth]{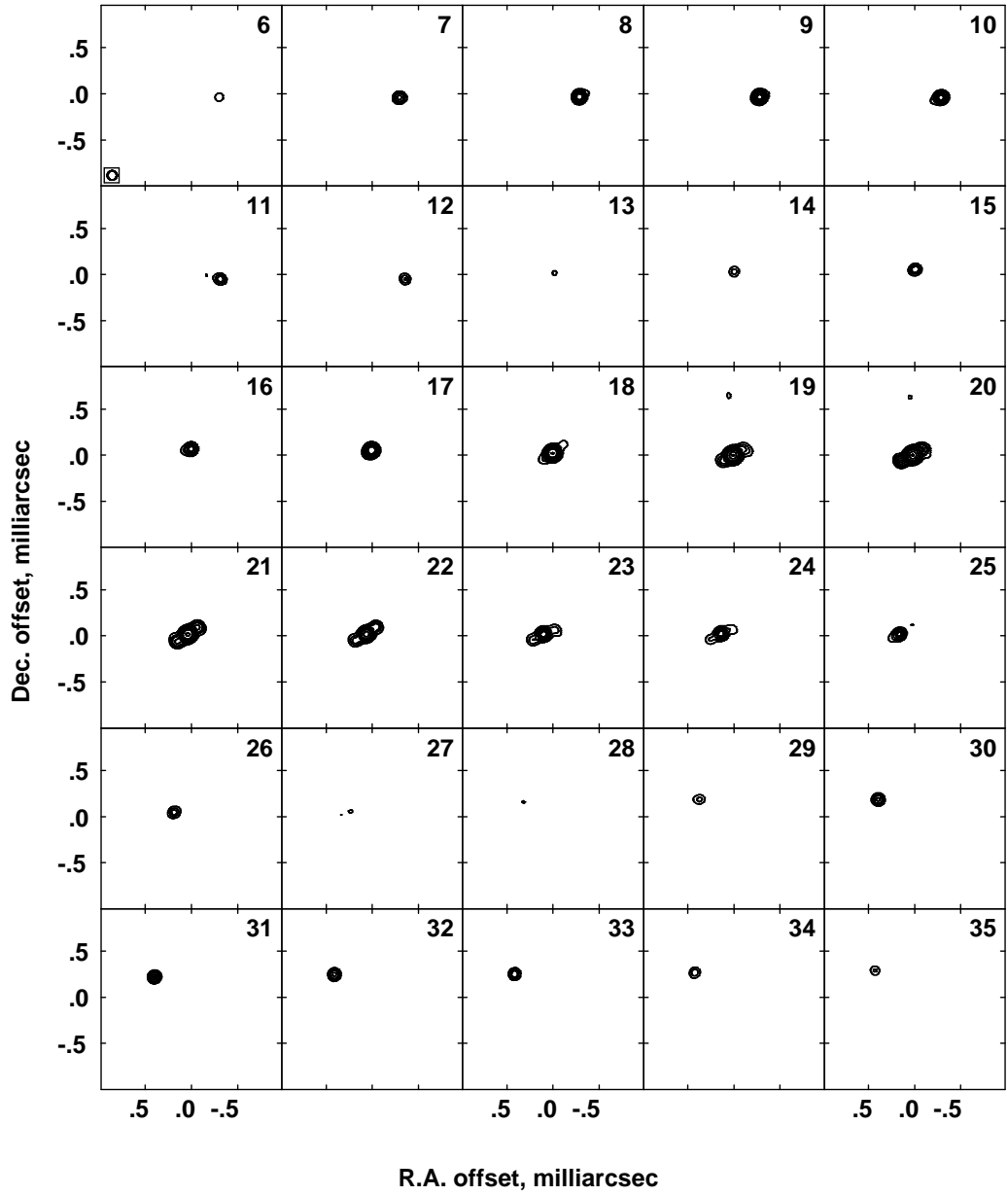}
\vskip -1.5cm
\caption{Images of the southern group of water maser spots in various
channels. The contours are drawn at $3.7\times(0.3, 0.5, 0.7, 1, 1.2,
1.5, 2, 3, 5, 6, 7, 8, 9)$~Jy/beam. \hfill}
\end{figure*}
\afterpage{\clearpage}

\begin{figure*}[t!]
\includegraphics[width=\linewidth]{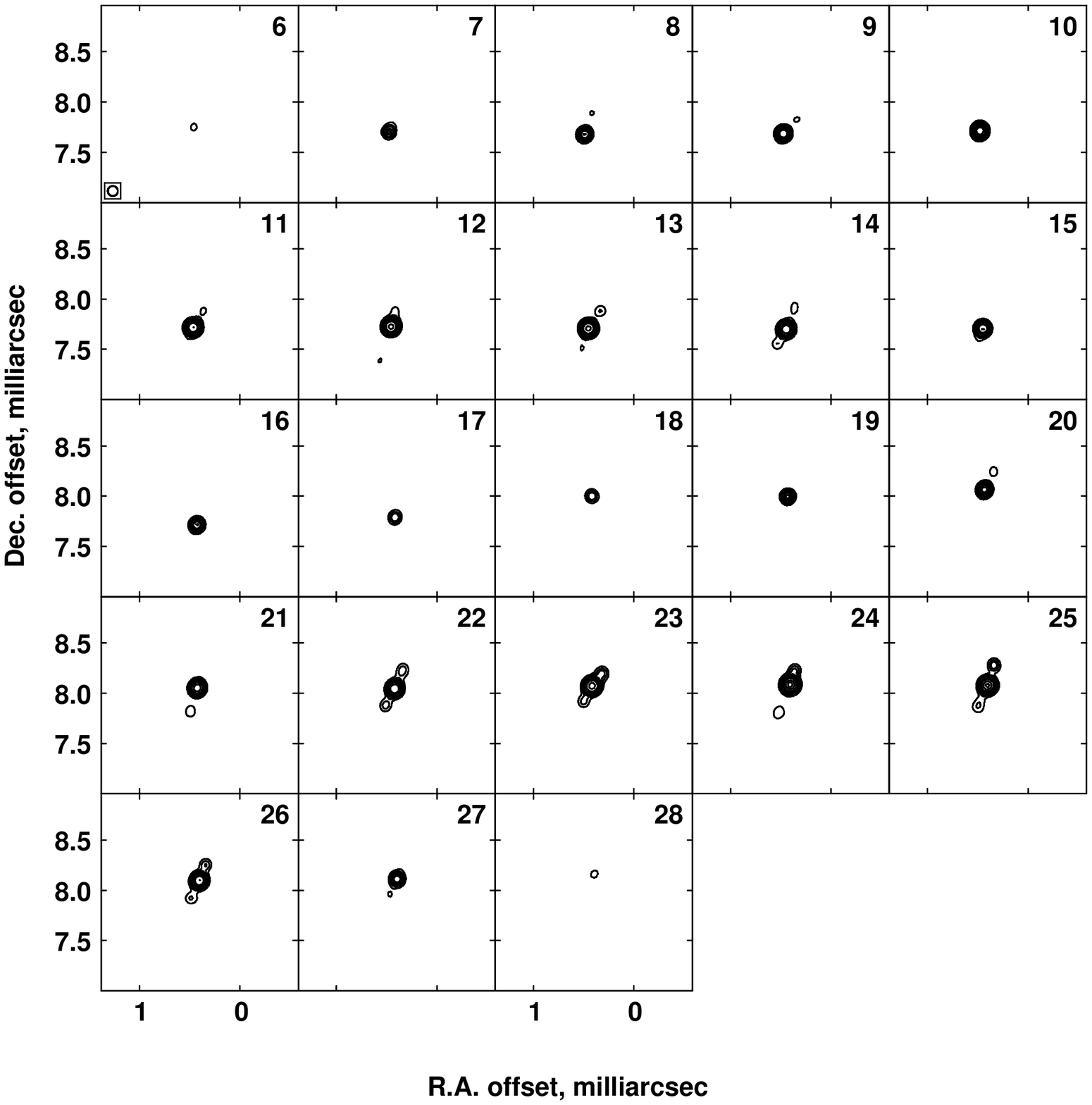}
\vskip -1cm
\caption{Images of the northern group of water maser spots in various
channels. The contours are drawn at $1.1\times(0.3, 0.5, 0.7, 1, 1.2,
1.5, 2, 3, 5, 6, 7, 8, 9)$~Jy/beam. \hfill}
\end{figure*}

\begin{figure*}[t!]
\vskip -1cm
\includegraphics[width=\linewidth]{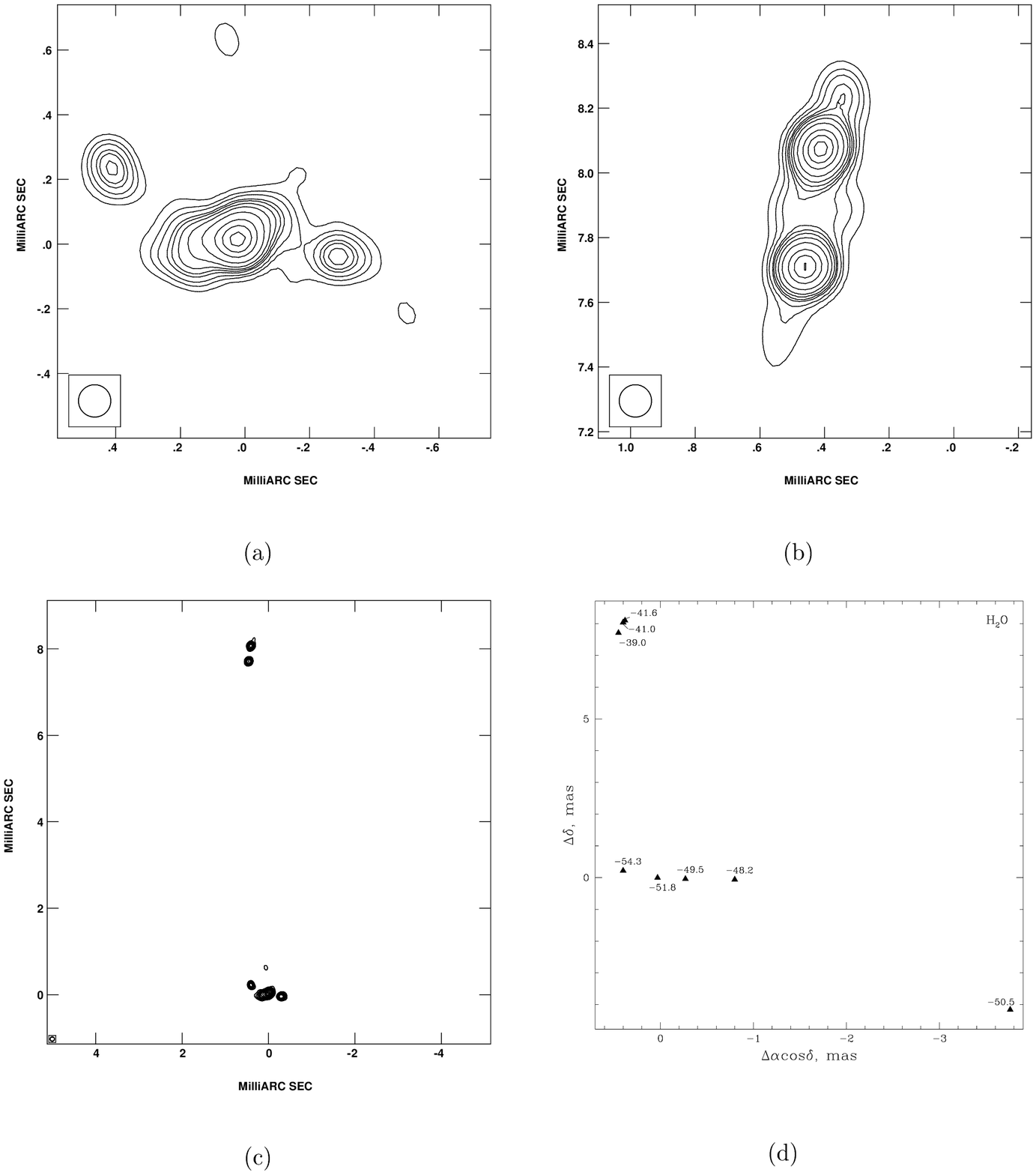}
\vskip -3mm
\caption{Summed maps of the H$_2$O masers of the (a) southern and (b)
northern groups. (c) Total map, where the brightness of each point is
determined by the maximum of the brightnesses of the maps at that point
for all spectral channels. The contours are drawn at 0.03, 0.05, 0.07, 0.1,
0.12, 0.15, 0.2, 0.3, 0.5, 0.6, 0.7, 0.8 and 0.9 of the maximum, which are
3.9~Jy/beam for map (a), 1.9~Jy/beam for map (b), and 3.9~Jy/beam for map
(c). (d) Positions of the water maser spots in GL~2789 relative to the
reference feature at~${-51.8}$~km/s. \hfill}
\end{figure*}

The spots themselves are very compact, and are not resolved even on the
longest VLBA baselines. Figure~6 shows the dependence of the correlated
flux on the baseline length for one of the spots. We can see that the
correlated flux is nearly independent of the baseline length, indicating
that this maser spot is not resolved by any of the baselines. The remaining
spots are also unresolved. The upper limit to their angular size is
0.05~mas (or 50~$\mu$arcsec), and the corresponding lower limit to their
brigthness temperature is $2\times10^{13}$~K.

\begin{figure*}[t!]
\vskip -1.3cm
\includegraphics[width=\linewidth]{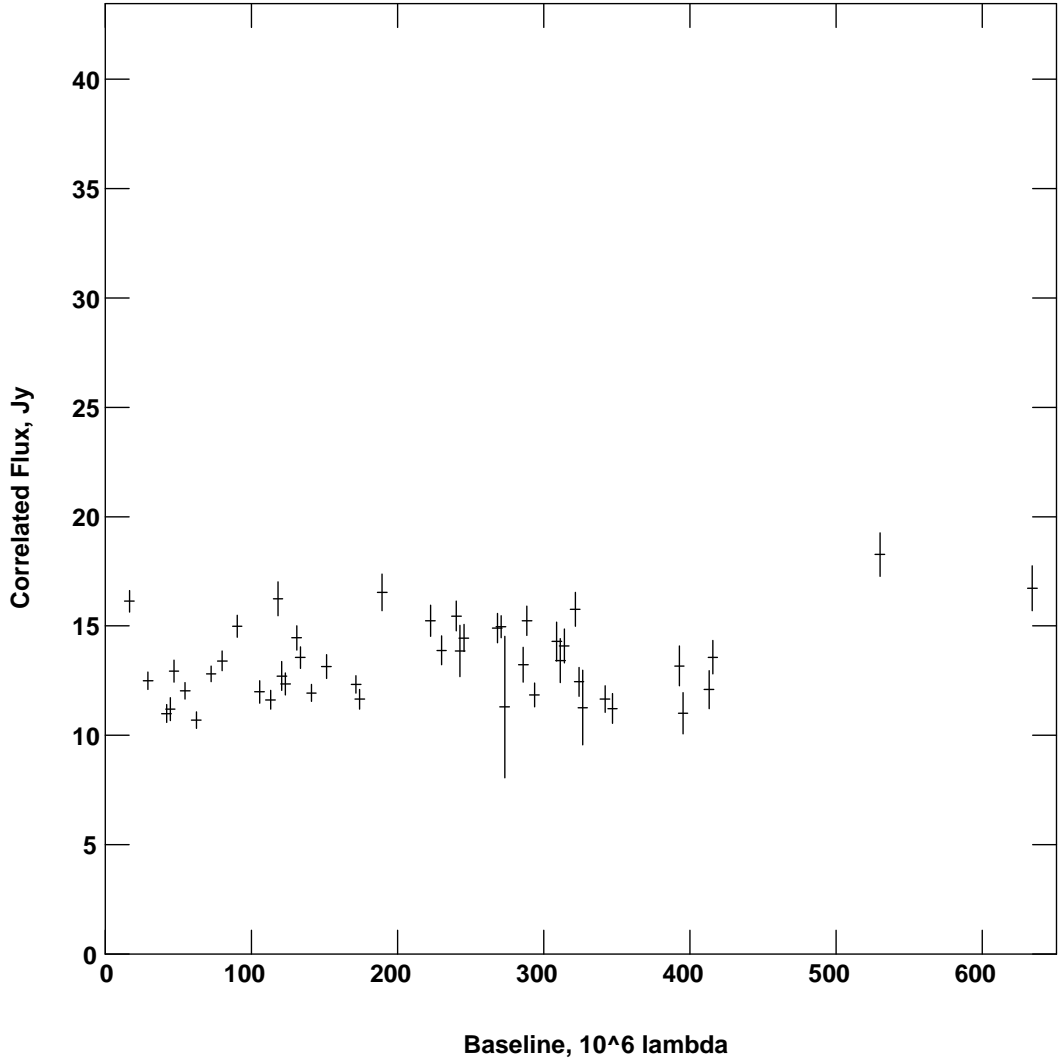}
\caption{Dependence of the correlated flux on the baseline length for one of
the water-maser spots (the feature at ${-41.6}$~km/s).\hfill}
\end{figure*}

\textbf{3.3.2. The Methanol maser.} The mapping of the methanol maser was
carried out in the same way as for the H$_2$O maser. The four spectral
features visible in Fig.~1 correspond to four maser spots (Table~3),
as we can see in Fig.~7a, which presents the image made from the 1998 data.
The spots are located roughly along a north--south line and form two
groups of two spots. The distance between the groups is about 100~mas,
and the radial velocity of the southern group is 2.5~km/s higher than that
of the northern group. As for the H$_2$O maser spots, the spots themselves
are not resolved; their sizes are less than 2~mas, corresponding to a lower
limit for the brightness temperature of $1.4\times10^9$~K. Figure~7b shows
the map of the GL~2789 methanol maser emission obtained using the 2000 data.
Since the methanol observations were carried out twice with a time interval
of two years, we can estimate possible changes in the relative positions
of the spots. A comparison of the two images indicates that the distances
between the spots did not change significantly, by less than 2~mas, which
is within the observational errors. The position of spot~$B$ changed by
2.5~mas in the line wings. The intensity and profile of the line corresponding
to spot~$B$ also varied. These variations may indicate a weakening of the
previous spot~$B$ and the appearance of another spot near the old position
at a slightly different velocity.

\begin{table}[b!]
\caption{Relative positions of the H$_2$O maser spots.}
\vskip 3mm
\begin{center}
\begin{tabular}{c|r|r|r}
\hline
\parbox[c]{2.5cm}{\strut Radial\\ velocity, km/s\strut}&
\multicolumn{1}{c|}{\parbox[c][1cm]{2cm}{$\Delta \alpha
\cos\delta $,\\ mas}}&
\multicolumn{1}{c|}{\parbox[c][1cm]{2cm}{$\Delta \delta $,\\ mas}}&\multicolumn{1}{c}{Flux, Jy}\\
\hline
$-$54.3 &     0.40 &    0.22 &  4.7 \\
$-$51.8 &     0.03 &    0.00 & 36.0 \\
$-$50.5 &  -3.76 & -4.16 &  1.8 \\
$-$49.5 &  -0.27 & -0.04 & 10.0 \\
$-$48.2 &  -0.80 & -0.06 &  0.6 \\
$-$41.6 &     0.38 &    8.10 &  6.6 \\
$-$41.0 &     0.40 &    8.04 &  6.4 \\
$-$39.0 &     0.45 &    7.71 &  7.3 \\
\hline
\end{tabular}
\end{center}
\end{table}

\section{MODEL FOR THE MASERS}

Figure~7c shows the relative positions of the H$_2$O and methanol masers.
In both masers, the spots are distributed roughly along a north--south line,
with the center of the methanol maser shifted 90~mas south from the center
of the H$_2$O maser. The accuracy of the position of the methanol maser
relative to the water maser is not high, so that it is reasonable to suppose
that both masers are located approximately on a north--south line similar
to those corresponding to the structure of the maser spots. The maximum
distance between the methanol-maser components is roughly a factor of ten
greater than for the H$_2$O maser, while the difference in radial velocities
is about a factor of three smaller. The radial velocity of the molecular
cloud associated with GL~2789, ${-45.6}$~km/s, is within the interval of
velocities for the H$_2$O maser and at the blue edge of the velocity interval
for the methanol maser. The radial-velocity gradients for the methanol and
H$_2$O masers have different signs: while the methanol radial velocity
decreases in the direction of the H$_2$O maser, from south to north, the
H$_2$O radial velocity increases from south to north.

\begin{table*}[t!]
\caption{Relative positions of the CH$_3$OH maser spots.}
\vskip 1mm
\begin{center}
\parbox{0.95\linewidth}{\footnotesize
The positions are given relative to the reference
water-maser spot at $-51.8$~km/s. This shift was determined from the
absolute-coordinate measurements and has an accuracy of about 50~mas. The
accuracy of the relative coordinates of the spots for lines emitted by
a single molecule is about 1~mas.\hfill}
\vskip 3mm
\begin{tabular}{|c|c|r|r|r|r|r|r|}
\hline
\multicolumn{2}{|c|}{}&\multicolumn{3}{|c|}{1998 Observations}&\multicolumn{3}{|c|}{2000 Observations}\\
\hline
Spectral&\cthead{Radial}&\cthead{$\Delta\alpha cos\delta$}&\cthead{$\Delta\delta$}&\cthead{Flux}&\cthead{$\Delta\alpha cos\delta$}&\cthead{$\Delta\delta$}&\cthead{Flux}\\
~feature~&\cthead{velocity}&\cthead{}&\cthead{}&\cthead{}&\cthead{}&\cthead{}&\cthead{}\\
&\cthead{km s$^{-1}$}&\cthead{mas}&\cthead{mas}&\cthead{Jy}&\cthead{mas}&\cthead{mas}&\cthead{Jy}\\
\hline
A&$-$43.4 &  76.0 & $-$45.1 &  2.0 & 77.7 & $-$45.4 &  2.5 \\
B&$-$42.5 &  85.0 & $-$65.0 &  2.5 &  83.5 & $-$64.4 &  1.2 \\
B\makebox[0pt]{$_{~\;1}^{~~*}$}&$-$42.8&83.7&$-$63.2& 1.0& 82.8 & $-$64.7 &  0.9 \\
C&$-$40.8 &  82.9 & $-$134.2 &  2.0 & 82.8 & $-$134.5 &  1.7 \\
D&$-$40.4 &  94.2 & $-$143.2 &  2.2 & 94.1 & $-$143.3 &  3.0 \\
\hline 
\end{tabular}
\end{center}
\vskip 1mm
{\footnotesize $^{*}$~component corresponds to the wing of the
feature at $-$42.5~km/s. \hfill}
\end{table*}

The variations of the radial velocity in the north--south direction are
shown in Fig.~7d. These properties can be explained using a model in which
the masers lie in a Keplerian disk viewed nearly edge-on rotating around
the supposed central star. The position of the star, which may correspond
to the optical object N0, is not known with sufficient accuracy to establish
the parameters of the Keplerian orbit. As we can see from Table~1, both
masers coincide with the object N0 to within the accuracy of its coordinates,
0\farcs2. From general considerations, it seems likely that the
H$_2$O masers are closer to the star than the methanol maser. This is supported
by their large radial-velocity dispersion, consistent with the idea that the
H$_2$O masers move in Keplerian orbits at closer distances to the star, and
the smaller linear scale of the H$_2$O maser source, about 50~AU (while the
size of the region of methanol-maser emission is of the order of 600~AU). In
addition, the H$_2$O maser requires a higher gas temperature
(\mbox{300--400}~K) and higher density ($\sim 10^9$~cm$^{-3}$), while the
methanol masers require temperatures not exceeding 100~ë and gas densities
not exceeding $10^7$~cm$^{-3}$.

It is natural to suppose that the gas temperature and density are higher
closer to the star than further from it. The opposite velocity gradients
for the H$_2$O and methanol masers are consistent with a disk model only
if the star is located between the northern and southern components of
the H$_2$O maser. In that case, the opposite radial-velocity gradient for
the H$_2$O maser can be explained by the fact that the masers are distributed
in the direction toward the star and have radial velocities that are close
to the systematic radial velocity of the star. The methanol masers arise
in directions tangential to the disk and have radial velocities equal to the
velocity of rotation of the disk at the corresponding distance from the
star. In GL~2789, we see only emission from the side of the disk that is
moving away from the observer and experiences a red shift. Components arising
in regions of the disk that are approaching and have blue shifts are not
visible for some reason, possibly due to differences in the pumping conditions.

Figure~8 shows a schematic of the disk and the distribution of the masers
and star. In this model, if the mass of the star is 10~$M_\odot$, the H$_2$O
masers are located at a distance ${R = 41.1}$~AU, while the methanol masers
are at distances between 200~AU and 800~AU, for the adopted distance to the
source of 6~kpc. This follows from the relation for a Keplerian disk~[13]
\begin{equation}
\frac {d V}{d x}=30 \textrm{~km~s}^{-1}\textrm{AU}^{-1}
\sqrt{M/R^3}, \label{one:Slysh_n}
\end{equation}
where $\Delta V$ is the observed radial-velocity interval ($\Delta V =
17.3$~km/s for the H$_2$O maser),  $\Delta È$ is the apparent extent of
the maser source ($\Delta È = 8$~mas (or 48~AU), $M$ is the mass of the
star in solar masses, and $R$ is the distance from the star to the H$_2$O
maser in AU. This model is similar to the model for a disk associated with a
black hole in the nucleus of the galaxy NGC~4258 [14]. In that case,
H$_2$O masers observed at radial velocities close to the systematic
velocity of the galaxy are distributed near the direction toward the
center of the disk, with high-velocity features located at tangential
points of the disk. In our model, in contrast to the model for NGC~4258,
the H$_2$O masers are observed only in the direction toward the center of
the disk, and the role of the high-velocity features is played by the
methanol masers, which show only blue features.

The disk model for the H$_2$O and methanol masers in GL~2789 can be verified
if time variations in the radial velocities of the H$_2$O components can be
detected: in the model, the expected acceleration is 1.1~km/s per year, and
the proper motion of the H$_2$O masers spots should be 0.52~mas/yr. Both
effects are associated with rotation of the disk, whose linear velocity
is 14.8~km/s at a distance of 41.1~AU from a star with a mass of 10~$M_\odot$,
where the H$_2$O masers are located. The acceleration and proper motion
of the methanol masers should be small, since they are observed in the
tangential direction. The disk model predicts a Keplerian dependence of
the radial velocity of the methanol masers on their distance from the
star, $V\sim R^{-\frac 1 2}$. The results of observations shown in Fig.~7d
are consistent with the Keplerian model, although the number of points is
too small to draw definitive conclusions. No proper motions of the methanol
maser spots relative to the reference feature were detected, as was noted
above, and an upper limit to the proper motion of 2~mas over 2~yrs
corresponds to an upper limit for the linear velocity of the relative motion
of the maser spots of 28.5~km/s. This limit is consistent with the absence
of proper motion during tangential motion of the maser spots.

\begin{figure*}[t!]
\vskip -0.6cm
\includegraphics[width=\linewidth]{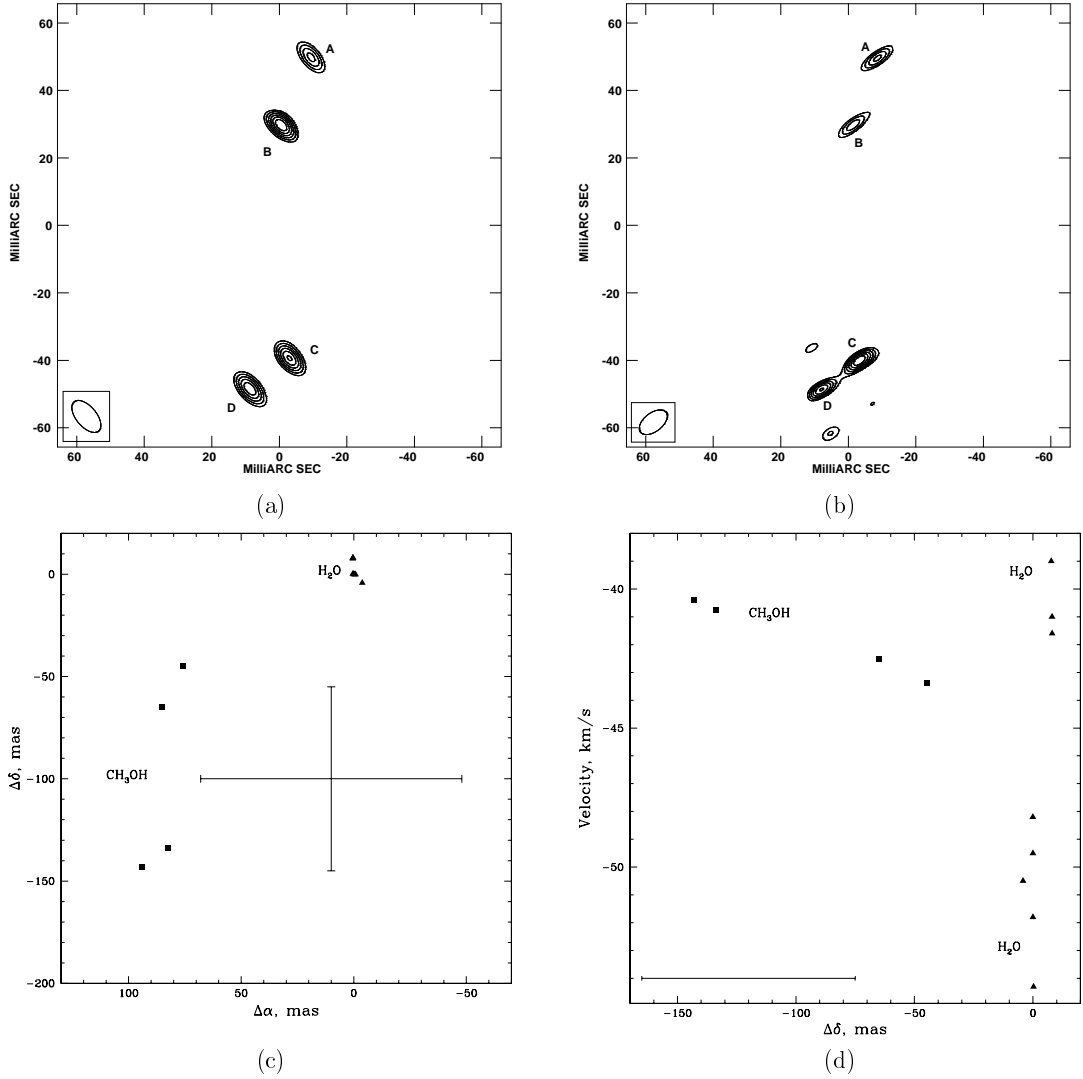}
\caption{Map of the methanol maser in GL~2789. (a) 1998 observations
with contours at $0.22\times(4, 5, 6, 7, 8, 9)$~Jy/beam. (b) 2000
observations with contours at $0.35\times(4, 5, 6, 7, 8, 9)$~Jy/beam.
Components A, B, C, and D correspond to spectral features with velocities
of ${-43.4}$, ${-42.5}$, ${-40.7}$, and ${-40.4}$, respectively. (c) Relative
positions of the methanol and water maser spots. The cross shows the error
in the relative position of the methanol and water masers. (d) Variations
of the radial velocity in the north--south direction. The error in the
relative position of H$_2$O and CH$_3$OH masers is shown in the lower left
corner.\hfill}
\end{figure*}

We can also consider other models for the GL~2789 maser sources. Historically,
the first models for H$_2$O masers associated with massive stars were
expanding-envelope models~[15]. This type of model is also of interest for
GL~2789, since it is known from optical observations that the star emits
a stellar wind with a velocity of more than 600~km/s [1]. However, such
high velocities are not observed in the H$_2$O and methanol maser lines.
Torrelles et al.~[16] suggest H$_2$O masers as associated with one-sided
jets ejected by the star. Such a jet is observed in the radio continuum of
W75N~[16]. However, no jets have been detected in the case of GL~2789; in
addition, the radio continuum source is much larger than the
maser-emission region~[6], and is probably associated with the region of
the stellar wind that is traced by the SII optical lines. The dimensions
of the region of SII emission and of the region of radio-continuum emission
coincide, as do the estimates of the mean density in this zone derived from
these observations, of the order of $2\times10^4$~cm$^{-3}$. The analysis
of the SII line profile performed by Hamann and Persson~[17] led them to
conclude that, in addition to its stellar wind, V645~Cyg possesses a dense
circumstellar disk. This was based on the presence of only blue components
in the SII line profile, corresponding to material moving toward the
observer. The red wing associated with the stellar-wind material moving
away from the observer was absent from the SII spectrum, and Hamann and
Persson~[17] proposed that the emission in the red wing was screened by
the disk. The presence of a disk is also supported by the emission in the
KI, FeI, and CaII lines, which arise in comparatively cool material
with a density of the order of $10^{10}$~cm$^{-3}$.

\begin{figure*}[t!]
\includegraphics[width=\linewidth]{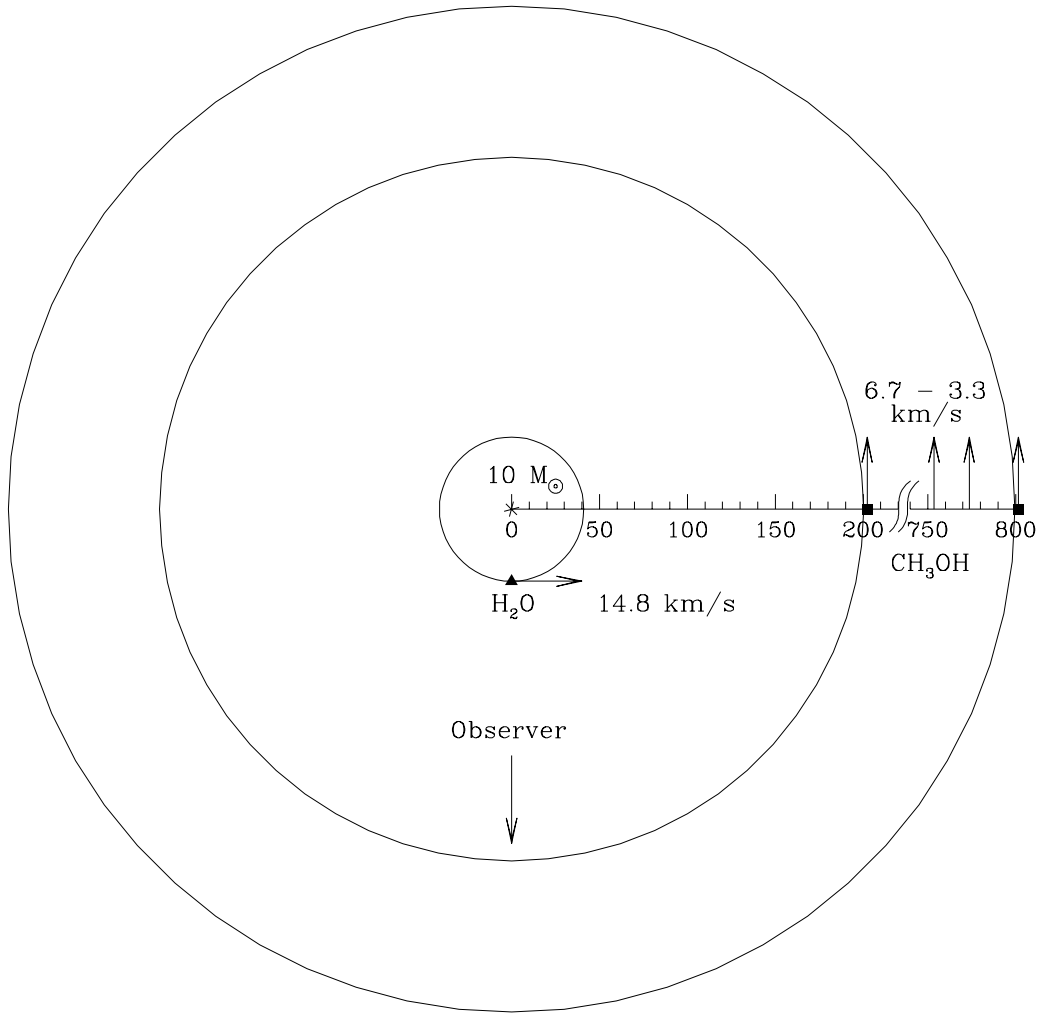}
\caption{Schematic of the disk in GL~2789. The numbers show distance from
the star in AU, and the arrows show the direction and magnitude of the
orbital velocity in km/s for the H$_2$O and methanol masers. \hfill}
\end{figure*}

Figure~6 of~[17] shows a model with a disk surrounding the star, together
with an extended region of stellar wind radiating in SII and H$_\alpha$
lines. In this model, the H$_2$O and methanol masers could be located in
the disk proposed by Hamman and Persson~[17]. Since the densities required
for the H$_2$O, and especially the methanol, masers are lower than the
density of the inner part of the disk derived from optical line data,
these masers must be located in a lower-density region of the disk,
at larger distances from the star.

In this model, the bipolar flow observed in GL~2789~[4] is emitted
perpendicular to the disk and moves away from the star and masers at
very large distances or tens of thousands of AU.

\section{DISCUSSION}

The detection of a class~II methanol maser and the demonstration that
it is associated with the optical object V645~Cyg, as well as the confirmation
of the connection between this object and the H$_2$O maser, provide
additional arguments that V645~Cyg is a young stellar object. The source
of thermal radio emission surrounding this object is most likely not a
classical HII region, and instead a region filled with stellar wind.

Another important characteristic of this young stellar object is the presence
of a cool circumstellar disk, in which compact sources of H$_2$O, methanol,
and possibly OH maser emission are immersed. In~[18], which was dedicated to
the methanol masers in W3(OH), it was suggested that individual features
in the methanol and OH spectra were generated in extended envelopes surrounding
ice planets. It is natural to suppose that such planets are located in a
disk and move in Keplerian orbits. The chemical composition of these
extended envelopes reflects the composition of the interstellar dust grains,
or more precisely of their mantles. It is known that the mantles of
dust grains consist primarily of ice, and methanol is the second most
abundant ice material after water. For example, in the direction toward
the high-mass protostars GL7009S and W33A, the abundances of methanol
ice relative to water ice are $30\%$ and $5-22\%$ for GL7009S and
W33A, respectively~[19]. The methanol abundance in the direction of NGC~7538
IRS9 is 25$\%$~[20], and is $8.7\%$ in the direction of GL~2136~[21]. In
other sources that have been studied, the methanol abundances does not
exceed several per cent~[19].

In cool disks surrounding protostars and very young stars, dust grains
can adhere to larger bodies, ultimately reaching the dimensions of small
planets. This is how the objects in Kuiper's belt were formed in the solar
system. At sufficiently close distances from the star, water and methanol
evaporate from the surfaces of planets and leave the gravitational fields of
these planets if their masses are sufficiently low. Envelopes consisting
primarily of water and methanol will thus form around these planets, whose
extent can be much larger than the dimensions of a planet. This same picture
is observed in miniature in our solar system, with cometary nuclei in place
of ice planets. With approach toward the Sun, ice and methanol are
evaporated from the surfaces of cometary nuclei, forming extended envelopes.

The water in such envelopes dissociates into atoms of hydrogen and
molecules of OH, and conditions favorable for water, OH, and methanol
maser emission are created: (a) rather high abundances, reaching
90\% for water and OH and about 10\% for methanol~"--- much higher than in
the interstellar medium; (b) high column densities of water and methanol
molecules~"--- $10^{16}$~cm$^{-2}$ for methanol and $10^{18}$~cm$^{-2}$ for
H$_2$O for an envelope $10^9$~cm (10\,000~km) in size~"---
required to create high optical depths and, accordingly, high maser
amplifications; (c) the required ranges of temperature and density~"---
from 400~K and $10^9$~cm$^{-3}$ for the H$_2$O masers to 100~K and
$10^6-10^7$~cm$^{-3}$ for the OH and class~II methanol masers;
(d) intense radiation near the rotational transitions of these molecules
(submillimeter wavelengths and the far infrared), provided by reprocessing
of the radiation of the young star by the dust.

These conditions are adopted in standard models for maser pumping. It is
obvious that comets in the solar system are too small to form such extended
dense envelopes; the optical depth in lines of H$_2$O and methanol is too
low, and no maser emission is observed. In GL~2789, the planets responsible
for the water-vapor line emission move in Keplerian orbits at a distance
from the star of about 40~AU, with each of the eight maser spots (Table~2)
corresponding to a separate planet. The OH and methanol masers are located
further from the star, at distances of 200--800~AU. At such large distances,
the orbits of planets may not be located in a disk, due to possible
perturbations in their motions. This could explain the presence of individual
maser spots outside the disk.

This simple model can easily be verified using further observations. This
requires monitoring of the radial velocities and positions of the maser
spots to measure accelerations and linear translations due to the Keplerian
motion. The most suitable masers for such observations are the H$_2$O masers,
which should be in the orbits that are closest to the star and move with
the highest speeds. The corresponding estimates are presented in the previous
section. In the model presented here, the period of revolution of the planets
responsible for the H$_2$O maser emission is about 83~yrs. Therefore, using
spectral and angular measurements with sufficiently high accuracy, it should
be possible to determine the accelerations and velocities of the maser spots
over fairly short time scales, less than one year. Such measurements must
be conducted often due to the strong variability of the H$_2$O maser
emission. However, on short time scales, the accuracy of measurements
such as proper-motion measurements could prove to be insufficient, even
using arrays with the longest possible baselines, such as the VLBA.

One possibility is the use of ground--space interferometers, which can
attain angular resolutions exceeding that of ground-based interferometers
by a factor of a few, or even an order of magnitude more than this. The
measurement of proper motions using ground--space interferometers requires
that the masers have sufficiently small angular sizes~"--- smaller than the
width of the interferometer fringe. We have shown that the maser spots in
GL~2789 are very compact, and are not resolved by the VLBA (Fig.~2b, Fig.~6).
Therefore, they may be suitable for measurements using a ground--space
interferometer. Other models with expanding components or jets predict
that the spots should move uniformly and linearly, which could also be
verified using observations with very high angular resolution.

\section{CONCLUSION}

The linear structure of the H$_2$O and methanol masers is in best agreement
with a disk model. If the mass of the star is 10~$M_\odot$ and it is located
at a distance of 6~kpc, the H$_2$O maser components are located about
40~AU from the star, with the methanol masers at distances of 200--800~AU.
The physical objects responsible for the emission of individual
maser spectral features could be ice planets surrounded by extended
water--methanol envelopes moving in Keplerian orbits around the star.
Such planets are similar to the objects of Kuiper's belt in the solar
system. This model can be verified using observations with high angular
resolution, using either ground-based or ground--space interferometers.

\section{ACKNOWLEDGEMENTS}

The authors affiliated with the Astro Space Center (VIS, MAV, and
IEV) carried out this work in the framework of the RadioAstron project
with partial support from the Russian Foundation for Basic Research
(project code~01-02-16902), INTAS (grant~97-11451) and the State Centers
for Science and Technology program "Astronomy" (grant~1.3.4.2).
The observational data
presented here were obtained using the VLBA of the National Radio
Astronomy Observatory (USA), which is supported by the National Science
Foundation and is operated under contract to Associated Universities, Inc.,
and also on the European VLBI Network, which is a consortium of European
and Chinese radio astronomy institutes supported by the various national
science councils.

Translated by D.~Gabuzda
\end{document}